\documentclass[12pt,prb,aps]{revtex4-1}
\usepackage{amsmath}           		          	
\usepackage{graphicx,epstopdf}					
\usepackage{amssymb}
\usepackage{fullpage}
\usepackage{color}
\usepackage{esint}
\pdfoutput = 1 

\allowdisplaybreaks

\begin{document}
\title{A Simple Model of Current Ramp-Up and Ramp-Down in Tokamaks}
\author{Richard Fitzpatrick\,\footnote{rfitzp@utexas.edu}}
\affiliation{Institute for Fusion Studies, Department of Physics, University of Texas at Austin, Austin TX 78712}

\begin{abstract}
A simple model of the ramp-up and ramp-down of the toroidal current in a tokamak plasma is developed. Faraday's law of electric induction
is found to limit how rapidly the current can be safety ramped up or down. It is estimated that the minimum completely safe ramp-up/down times for the JET, SPARC, and
ITER tokamaks are 4.2, 2.0, and 14.7 seconds, respectively. The JET ramp time is in accordance with operational experience. The SPARC
and ITER minimum safe ramp times are less than the ramp times in the respective designs. Hence, there is no indication that the design ramp times are infeasible, as was recently 
suggested  in arXiv:2507.05456 (2025).
The typical ratios of the inductive electric field
to the Connor-Hastie field in SPARC and ITER are found to be less than those in JET. Thus, the fact that the JET tokamak was able to operate successfully
without encountering runaway electron problems during current ramps suggests that the future SPARC and ITER tokamaks should also be able to
avoid such problems. 
\end{abstract}
\maketitle

\section{Introduction}
In a recent paper, Boozer\,\cite{boozer} argues  that Faraday's law of electric induction places strong constraints on the operation of a future  tokamak fusion power plant.
In particular, Boozer takes issue with the claim by de Vries et alia (2018)\,\cite{deVries} that a fully controlled current ramp-down in the future ITER tokamak, starting from a toroidal plasma current of $15$ MA,
can be effected in 60 seconds. (Indeed, in an earlier study Boozer\,\cite{boozer1} seems to argue that the true ramp time is 1000 seconds.)  Boozer also takes issue with the claim by Creely et alia\,\cite{creely} that the plasma current in the future SPARC tokamak can be ramped up in
7 seconds, and ramped down in 12 seconds, and seems to imply that the true ramp time is, in fact, 1157 seconds. The claim by de Vries et alia (2018) is
made in Sect.~2 of Ref.~\onlinecite{deVries}, and is based on simulations performed using  the CORSICA\,\cite{corsica,corsica1}  code. The claim by Creely et alia is deduced from 
in Fig.~3 of Ref.~\onlinecite{creely}, and is based on simulations made with the TSC\,\cite{tsc}   code. While it is the case that the  CORSICA and TSC codes both have a proven 
track records, we can certainly agree with Boozer that neither de Vries et alia (2018) or Creely et alia provide sufficient information to validate their claims. 
To help clarify the situation, this paper presents a simple model of current ramp-up and ramp-down in tokamak plasmas. 
The aim of the paper is
to determine, in a transparent manner, whether the claims by de Vries et alia (2018) and Creely et alia are, at least, plausible, and also to clarify the constraints that Faraday's law
actually do place on future tokamak operation. Boozer\,\cite{boozer,boozer1}  also highlights the danger
of runaway electron generation during current ramps, so we shall additionally try to estimate the likelihood of dangerous levels of  runaway electron generation during current ramps
in  future tokamaks.  

\section{Tokamak Operation}
\begin{figure}
\centerline{\includegraphics[width=0.85\textwidth]{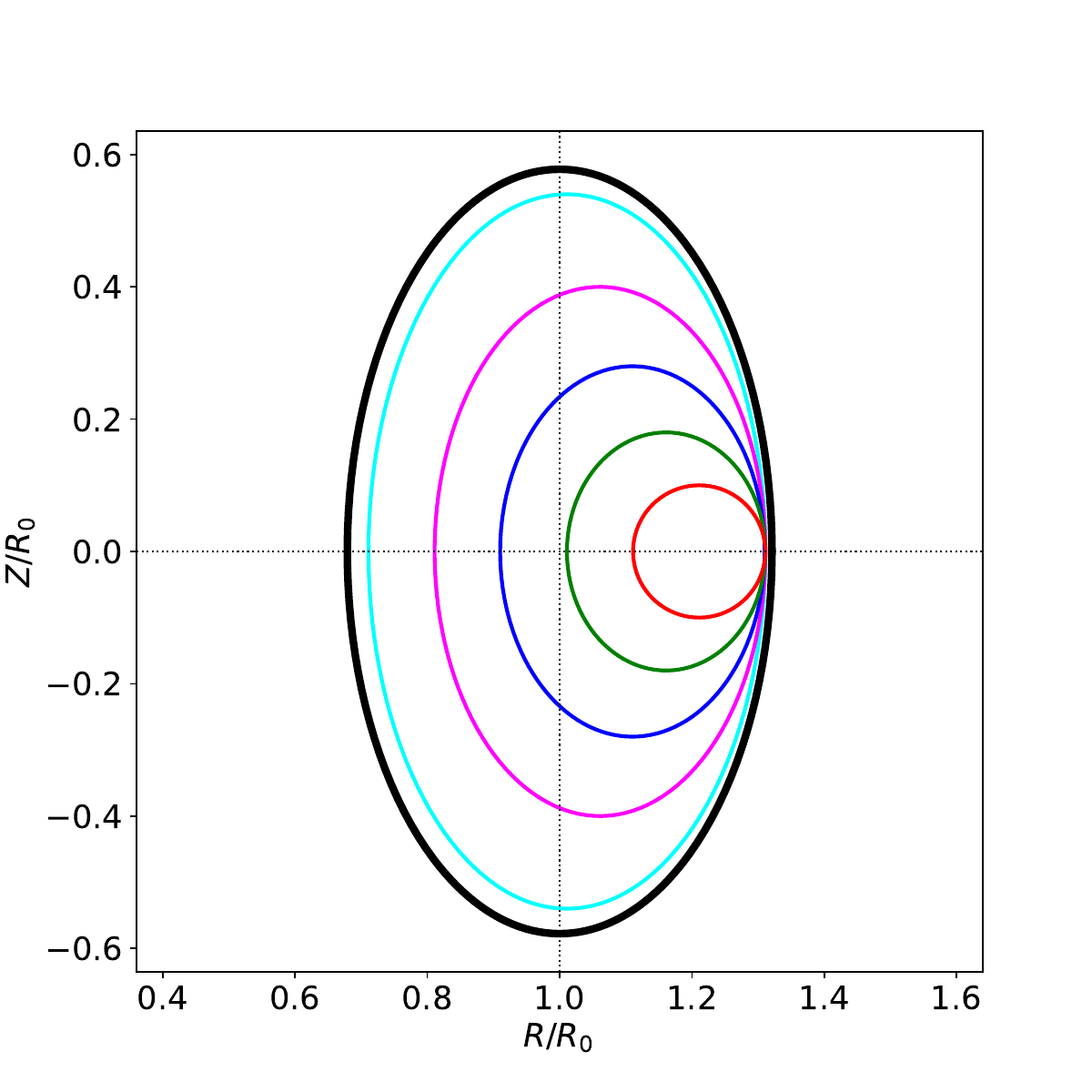}}
\caption{A cartoon showing the poloidal cross-section of a conventional tokamak during a current ramp-up. The black curve represents the first wall. The
red, green, blue, magenta, and cyan curves show, in order of increasing time, the plasma boundary at various stages of the current ramp-up.
Here, $R$, $\phi$, $Z$ are conventional cylindrical coordinates that are co-axial with the plasma torus. }\label{fig1}
\end{figure}

Figure~\ref{fig1} is a cartoon that illustrates how a toroidal current ramp-up is effected in a conventional tokamak plasma.\cite{lister,federici, jackson,politzer}
After the initial burn-through phase, the fully ionized plasma starts off as an {\em ohmic}\/  discharge of circular poloidal cross-section, and relatively
small minor radius, that is limited on the first wall, and carries a comparatively small toroidal current. As the toroidal current is
gradually increased, the minor radius and vertical elongation of the plasma increase in concert, but the plasma remains limited on
the first wall. (See Fig.~7 of Ref.~\onlinecite{lister} and Fig.~1 of Ref.~\onlinecite{federici}.) Only in the very final stages of the current ramp is the magnetic X-point added to the equilibrium, at which point the
plasma becomes magnetically diverted, and consequently ceases to be limited by the first wall. 

The current ramp usually takes place in two stages.  (See Fig.~3 of Ref.~\onlinecite{creely}.) In the first stage, the edge safety-factor, $q_a$, of the  plasma is allowed to decrease from
an initially very large value to its final value of about 3.5, at fixed plasma minor radius. In the second stage, which is usually much longer in duration than the first, $q_a$ is held constant
while the plasma expands in volume. The first stage is hazardous because, as $q_a$ passes through integer values (6, 5, 4, et cetera), the plasma becomes
momentarily susceptible  to both tearing and ideal-kink modes.\cite{wesson,granetz1,cheng} However, the degree of the hazard is mitigated by the smallness of the plasma current.
In the second stage, the safety-factor profile is  held approximately constant, and is designed to be stable to both tearing and ideal-kink modes. 

After the current ramp has been completed, and the magnetic diverter is operational,  external heating is applied to the plasma discharge, while the  current is held constant.
Consequently, 
the plasma eventually attains its peak temperature. The current flat-top phase of the plasma discharge is usually much longer in duration than both the current ramp-up and ramp-down
phases. Toward the end of the current flat-top, the external heating is turned off, and the plasma eventually returns to the same state that it was in at the end of the
current ramp-up. At this point, the sequence of events that constitute the current ramp-up is carried out in reverse order to effect a current ramp-down. In other words, the magnetic X-point
is first removed from the plasma, which  then becomes limited by the first wall. The plasma is then gradually crushed against the first wall, while the current is reduced, but
$q_a$ is held constant. (See Fig.~7 of Ref.~\onlinecite{lister}.) Finally, $q_a$ is allowed to rise to a large value, at fixed minor radius,  as last remnants of the plasma current disappear. 

\section{Analysis}
\subsection{Coordinates}
Let us model a tokamak plasma, in the simplest imaginable manner, as a periodic cylinder. Let $r$, $\theta$, $z$ be conventional cylindrical coordinates, and
let the magnetic axis of the plasma correspond to $r=0$. The system is assumed to be periodic in $z$ with period length $2\pi\,R_0$, where $R_0$ is the
simulated major radius of the plasma. Let $a$ be the minor radius of the plasma at the end of the current ramp. 

\subsection{Fundamental Equations}
The equilibrium magnetic field is represented as
\begin{equation}
{\bf B}(r,t) = B_\theta(r,t)\,{\bf e}_\theta+ B_0\,{\bf e}_z,
\end{equation}
where $B_0$ is the constant toroidal magnetic field generated by  currents flowing in external  field-coils. The poloidal field, $B_\theta(r,t)$, on the other hand, is generated
via transformer action by the time variation of the current flowing in the tokamak's primary winding. The current density in the plasma is written ${\bf j}= j_z\,{\bf e}_z$. Moreover, Amp\`{e}re's law implies that
\begin{equation}
\mu_0\,j_z(r,t) = \frac{1}{r}\,\frac{\partial}{\partial r}(r\,B_\theta).
\end{equation}
The total toroidal current carried by the plasma is
\begin{equation}
I_p(t)= \gamma_s\int_0^a 2\pi\,r\,j_z\,dr = \frac{\gamma_s\,2\pi\,a\,B_\theta(a,t)}{\mu_0},
\end{equation}
where
\begin{equation}
\gamma_s = \frac{1+\kappa^2}{2}
\end{equation}
is an empirically determined factor that specifies the increase in the toroidal current over the cylindrical value in a plasma of vertical elongation $\kappa$.\cite{creely,uckam}
The plasma safety-factor profile takes the form 
\begin{equation}
q(r,t) = \frac{r\,B_0}{R_0\,B_\theta},
\end{equation}
while the electric field in the plasma is written ${\bf E}= E_z\,{\bf e}_z$. Faraday's law yields
\begin{equation}
\frac{\partial E_z}{\partial r} = \frac{\partial B_\theta}{\partial t},
\end{equation}
while Ohm's law gives
\begin{equation}
E_z = \eta\,j_z,
\end{equation}
where
\begin{equation}
\eta = \frac{Z\,\ln{\mit\Lambda}}{1.96\,6\sqrt{2}\,\pi^{3/2}}\,\frac{m_e^{\,1/2}\,e^{\,2}\,c^{\,4}\,\mu_0^{\,2}}{T_e^{\,3/2}}
\end{equation}
is the Spitzer plasma resistivity.\cite{spitzer,fitz} Here, $T_e(r,t)$ is the electron temperature profile, $Z$ the effective ion charge number, $\ln{\mit\Lambda}\simeq 15$ the
Coulomb logarithm, $m_e$ the electron mass, $e$ the magnitude of the electron charge, $c$ the velocity of light in vacuum, and
$\mu_0$ the vacuum permeability. 
Note that, for the sake of simplicity, we are neglecting both the neoclassical enhancement of the plasma resistivity, as well as 
the non-inductive bootstrap current.\cite{fitz1} The neoclassical  enhancement could most easily be taken into account in our analysis by multiplying $Z$
 by the volume-averaged enhancement factor. 
 Finally, it is helpful to define the poloidal magnetic flux, $\psi(r,t)$, where 
\begin{equation}
B_\theta =  \frac{1}{2\pi\,R_0}\,\frac{\partial\psi}{\partial r}.
\end{equation}

The electron energy balance equation takes the form\,\cite{fitz}
\begin{equation}
\frac{3}{2}\,n_e\,\frac{\partial T_e}{\partial t} - n_e\,\frac{1}{r}\,\frac{\partial}{\partial r}\left(r\,\chi_\perp\,\frac{\partial T_e}{\partial r}\right)
 = \eta\,j_z^{\,2},
 \end{equation}
 where $n_e$ is the electron number density, and $\chi_\perp(r)$ is the electron perpendicular energy diffusivity due to small-scale plasma turbulence. 
 Note that, for the sake of simplicity, we are treating $n_e$ as a spatial and temporal constant. Moreover, we expect $\chi_\perp \sim 1\,{\rm m^2\,s}^{-1}$.\cite{book}
 Finally, the net power input to the plasma due to ohmic heating is
 \begin{equation}
 P(t) = \gamma_p\,4\pi^2\,R_0\int_0^a r\,\eta\,j_z^{\,2}\,dr,
 \end{equation}
 where $\gamma_p= (\gamma_s/\kappa)^2$ is an empirical factor that takes the shaping of the plasma poloidal cross-section into account.\cite{uckam}
 
 \subsection{Fundamental Quantities}
 Let $\epsilon= a/R_0$ be the inverse aspect-ratio of the plasma, $q_a$ the value of the edge safety-factor at the end of the current ramp, and 
 \begin{equation}
 B_{\theta\,a}= \frac{\epsilon}{q_a}\,B_0
 \end{equation}
 the edge poloidal magnetic field at the end of the current ramp. Suppose that $\chi_\perp (r)= \chi_0\,\hat{\chi}(r)$, where $\chi_0$ is a
 typical value of the perpendicular energy diffusivity, and $\hat{\chi}(r)$ is dimensionless and of order unity. 
 We can define
 \begin{align}
 T_0&= \left[\frac{\,Z\,\ln{\mit\Lambda}}{1.96\,6\sqrt{2}\,\pi^{3/2}}\,\frac{m_e^{\,1/2}\,e^{\,2}\,c^{\,4}\,B_{\theta\,a}^{\,2}}{n_e\,\chi_0}\right]^{2/5}\\[0.5ex]
 &= 9.95\times 10^{-1}\,Z^{\,2/5}\,{\mit\Lambda}_{15}^{\,2/5}\,n_{20}^{-2/5}\,\chi_0^{-2/5}\,B_{\theta\,a}^{\,4/5}\,\,[{\rm keV}]
 \end{align}
 as the typical electron temperature attained at the end of the current ramp. Here, ${\mit\Lambda}_{15}=\ln{\mit\Lambda}/15$ and $n_{20}=n_e/10^{\,20}$. 
 All other quantities are in SI units. Likewise, 
 \begin{equation}
 \tau_R \equiv \frac{a^2\,\mu_0}{\eta(T_0)}= 4.99\times 10^1\,a^2\,Z^{-2/5}\,{\mit\Lambda}_{15}^{-2/5}\,n_{20}^{-3/5}\,\chi_0^{-3/5}\,B_{\theta\,a}^{\,6/5}\,\,[{\rm s}]
 \end{equation}
 is the conventional resistive diffusion time, whereas 
 \begin{equation}
 \tau_c= \frac{a^2}{\chi_0}\,\,[{\rm s}]
 \end{equation}
 is the energy confinement time. The plasma poloidal beta is defined
 \begin{equation}\label{e16}
 \beta_p\equiv \frac{\mu_0\,n_e\,T_0}{B_{\theta\,a}^{\,2}} =  2.00\times 10^{-2}\,Z^{\,2/5}\,{\mit\Lambda}_{15}^{\,2/5}\,n_{20}^{3/5}\,\chi_0^{-2/5}\,B_{\theta\,a}^{-6/5}.
 \end{equation}
 Note that
 \begin{equation}\label{e17}
 \beta_p = \frac{\tau_c}{\tau_R}
 \end{equation}
 in an ohmic plasma. 
 The typical plasma current is
 \begin{equation}
 I_0= \frac{2\pi\,\gamma_s\,a\,B_{\theta\,a}}{\mu_0} = 5.00\,\gamma_s\,a\,B_{\theta\,a}\,\,[{\rm MA}],
 \end{equation}
 whereas typical inductive electric field-strength in the plasma takes the form
 \begin{equation}
 E_0 =\frac{\beta_p\,\chi_0\,B_{\theta\,a}}{a} = 2.00\times 10^{-2}\,a^{-1}\,Z^{\,2/5}\,{\mit\Lambda}_{15}^{\,2/5}\,n_{20}^{3/5}\,\chi_0^{\,3/5}\,B_{\theta\,a}^{-1/5}\,\,[{\rm V\,m}^{-1}],
 \end{equation}
 and the typical ohmic heating power is written
 \begin{equation}
 P_0 = 4\pi^2\,\gamma_p\,R_0\,n_e\,\chi_0\,T_0 = 6.29\times 10^{-1}\,\gamma_p\,R_0\,Z^{\,2/5}\,{\mit\Lambda}_{15}^{\,2/5}\,n_{20}^{3/5}\,\chi_0^{\,3/5}\,B_{\theta\,a}^{\,4/5}\,\,[{\rm MW}].
 \end{equation}
 Finally, the typical poloidal magnetic flux takes the form 
 \begin{equation}
 \psi_0= 6.28\,R_0\,a\,B_{\theta\,a}\,[{\rm V\,s}].
 \end{equation}
 
\subsection{Runaway Electron Generation}
 The plasma at the start of the current ramp is cold and resistive. Consequently, driving a current through the discharge requires a comparatively large inductive electric field. In such 
 circumstances, there is a danger of runaway electron generation. Runaway electrons are suprarthermal electrons for which the Coulomb collisional drag due to the bulk plasma is
 less than the acceleration due to the inductive electric field. Such electrons can acquire energies in excess of 10 MeV, and  cause considerable damage if they strike the 
 first wall. 
 
 Runaway electron generation is only possible, in theory,  when the electric field exceeds the so-called {\em Connor-Hastie}\/ value,\cite{connor}
 \begin{equation}
 E_c = \frac{\ln{\mit\Lambda}\,n_e\,e^{\,3}}{4\pi\,\epsilon_0^{\,2}\,m_e\,c^{\,2}} = 7.65\times 10^{-2}\,{\mit\Lambda}_{15}\,n_{20}\,\,[{\rm V\,m}^{-1}].
 \end{equation}
 Here, $\epsilon_0$ is the vacuum permittivity. 
 However, the criterion $E>E_c$ does not,  by itself, guarantee the generation of dangerous quantities of runaway electrons, because, before this
 can happen, there needs to be a source of suprathermal electrons. One obvious source is from the high-energy tail of the thermal electrons. However,
 it is well-known that this source does not become effective until the electric field approaches the so-called {\em Driecer}\/ value,\cite{dreicer}
 \begin{equation}
 E_D = E_c\,\frac{m_e\,c^{\,2}}{T_e}= E_{D\,0}\,\frac{T_e}{T_0},
 \end{equation}
 where
 \begin{equation}
 E_{D\,0} = 3.93\times 10^{1}\,Z^{-2/5}\,{\mit\Lambda}_{15}^{\,3/5}\,n_{20}^{7/5}\,\chi_0^{2/5}\,B_{\theta\,a}^{-4/5}\,\,[{\rm V\,m}^{-1}].
 \end{equation}
 Note that $E_{D\,0}\gg E_c$. 
 
 In an extensive experimental study of runaway electron generation in the JET tokamak,\cite{run} de Vries et alia (2020) find that JET
  plasmas invariably satisfy the runaway electron existence criterion, $E>E_c$, in the early stages of the discharge, but that this does not
 usually lead to the generation of dangerous quantities of runaway electrons. In fact, de Vries et alia (2020) conclude that $E>10\,E_c$ is
 needed before runaway electron generation becomes a problem. This conclusion is consistent with those of earlier experimental 
 studies.\cite{granetz,paz,pop}
 
\subsection{Normalization}
 Let us adopt the following convenient normalization scheme: $r=a\,\hat{r}$, $t=\tau_R\,\hat{t}$, $T_e=T_0\,\hat{T}(\hat{r},\hat{t})$, $B_\theta=B_{\theta\,a}\,\hat{B}_\theta(\hat{r},\hat{t})$,
 $I_p=I_0\,\hat{I}_p(\hat{t})$, $j_z=[B_{\theta\,a}/(\mu_0\,a)]\,\hat{j}$, $E=E_0\,\hat{E}(\hat{r},\hat{t})$, $E_c= E_0\,\hat{E}_c$, $E_D= E_0\,\hat{E}_D(\hat{r},\hat{t})$, $E_{D\,0}=E_0\,\hat{E}_{D\,0}$, $P= P_0\,\hat{P}(\hat{t})$,  and
 $\psi=\psi_0\,\hat{\psi}(\hat{r},\hat{t})$. It follows that 
 \begin{align}\label{e25}
\frac{3}{2}\,\beta_p\,\frac{\partial \hat{T}}{\partial\hat{t}}&=
 \frac{\partial^2\hat{T}}{\partial \hat{r}^2} 
 +\frac{1}{\hat{r}}\,\frac{\partial\hat{T}}{\partial\hat{r}} + \frac{d\ln\hat{\chi}}{d\hat{r}}\,\frac{\partial\hat{T}}{\partial\hat{r}}
 + \frac{\hat{E}^{\,2}\,\hat{T}^{\,3/2}}{\hat{\chi}},\\[0.5ex]
\frac{\partial\hat{B}_\theta}{\partial \hat{r}} + \frac{\hat{B}_\theta}{\hat{r}}&= \hat{E}\,\hat{T}^{\,3/2},\\[0.5ex]
\frac{\partial\hat{B}_\theta}{\partial\hat{t}}&= \frac{\partial\hat{E}}{\partial\hat{r}},\label{e27}
 \end{align}
 as well as
 \begin{align}
 \frac{q}{q_a}&= \frac{\hat{r}}{\hat{B}_\theta},\label{e28}\\[0.5ex]
 \hat{I}_p(\hat{r},\hat{t})&= \hat{r}\,\hat{B}_\theta(\hat{r},\hat{t}),\label{e29}\\[0.5ex]
 \hat{B}_\theta &=\frac{\partial\hat{\psi}}{\partial\hat{r}},\label{e30}\\[0.5ex]
 \hat{E}_D&= \hat{E}_{D\,0}\,\hat{T}^{-1},
 \end{align}
 Here, $\hat{I}_p(\hat{r},\hat{t})$ is the normalized toroidal plasma current contained within normalized minor radius $\hat{r}$. 
 Equation~(\ref{e27}) can be integrated to give
 \begin{equation}\label{e31}
 \hat{E}(\hat{r},\hat{t})= \frac{\partial\hat{\psi}}{\partial \hat{t}}={\cal E}(\hat{t})+\int_0^{\hat{r}}\,\frac{\partial\hat{B}_\theta}{\partial\hat{t}}\,d\hat{r}',
 \end{equation}
 where use has been made of Eq.~(\ref{e30}). It follows that 
 \begin{equation}
 {\cal E}(\hat{t}) = \frac{\partial\hat{\psi}(0,\hat{t})}{\partial\hat{t}}.
 \end{equation}
 Finally, it can be shown that
 \begin{align}
 \hat{E}_c&= 3.82\,a\,Z^{-2/5}\,{\mit\Lambda}_{15}^{\,3/5}\,n_{20}^{2/5}\,\chi_0^{-3/5}\,B_{\theta\,a}^{\,1/5},\\[0.5ex]
 \hat{E}_{D\,0} &= 1.96\times 10^3\,Z^{-4/5}\,{\mit\Lambda}_{15}^{\,3/5}\,n_{20}^{4/5}\,\chi_0^{-1/5}\,B_{\theta\,a}^{-3/5}.
 \end{align}
 
\subsection{Approximations}
In order to facilitate our analysis, we shall make two main approximations. 
Our first approximation is to neglect the term involving $\beta_p$ in Eq.~(\ref{e25}).
This approximation is warranted because, as is clear from Eq.~(\ref{e16}), $\beta_p$ is much less than unity in  conventional ohmic
tokamak plasmas. (See also Table~\ref{t2}.) As is apparent from Eq.~(\ref{e17}), $\beta_p$ is much less than unity because the energy confinement time is much less than
the resistive diffusion time in ohmic plasmas. The neglect of the term in question implies that, during the current ramp,  the temperature profile remains in a quasi-equilibrium state in which
the energy per unit time that flows across the plasma boundary, and is presumably absorbed by the limiter, matches the ohmic heating power. Under these
circumstances, the normalized ohmic heating power is written
\begin{equation}\label{e36}
\hat{P}(\hat{t}) = - \left(\hat{\chi}\,\hat{r}\,\frac{\partial\hat{T}}{\partial\hat{r}}\right)_{\hat{r}=1}.
\end{equation}

Our second approximation is to neglect the second term on the extreme right-hand side of Eq.~(\ref{e31}). The neglect of the term in question implies that
the inductive electric field is {\em uniform}\/ within the plasma. Of course, this is the usual situation during the current flat-top phase of a tokamak
discharge. However, it is possible for the electric field to be spatially uniform during a current ramp, provided that the current is not ramped up or down too
rapidly. Indeed, the criterion that must be satisfied is
\begin{equation}\label{e35}
{\cal E} >\left| \int_0^{1}\,\frac{\partial\hat{B}_\theta}{\partial\hat{t}}\,d\hat{r}'\right|.
\end{equation}

During the ramp-up phase, if the toroidal current is ramped up sufficiently rapidly that the criterion (\ref{e35}) is not satisfied then the electric field in the outer regions of the plasma becomes
greater than that in the core, leading to a broadening of the current profile. However, a broad current profile implies a low internal plasma inductance,
and low-inductance plasmas are prone to destructive ideal-kink instabilities.\cite{wesson,cheng} 
If the current is ramped-up extremely rapidly then the current profile becomes hollow, giving rise to
minor disruption events that rapidly relax the current profile toward a  more peaked profile, but also occasionally trigger major disruptions.\cite{granetz}
During the ramp-down phase, on the other hand, if the current is ramped down sufficiently rapidly that the criterion (\ref{e35}) is not satisfied then the electric field in the outer regions of the plasma becomes
less than that in the core, leading to a peaking of the current profile. However, plasmas with strongly peaked current profiles are prone to tearing instabilities
that can lead to major disruptions.\cite{wesson,cheng} Thus, the criterion (\ref{e35}) is a necessary one for  the completely safe operation of a tokamak during the
current ramp-up and ramp-down phases. 

Our two approximations lead to the following equations:
\begin{align}\label{e38}
\frac{\partial^2\hat{T}}{\partial \hat{r}^2} 
 +\frac{1}{\hat{r}}\,\frac{\partial\hat{T}}{\partial\hat{r}} + \frac{d\ln\hat{\chi}}{d\hat{r}}\,\frac{\partial\hat{T}}{\partial\hat{r}}
 + \frac{{\cal E}^{\,2}\,\hat{T}^{\,3/2}}{\hat{\chi}}&=0,\\[0.5ex]
 \frac{\partial\hat{B}_\theta}{\partial \hat{r}} + \frac{\hat{B}_\theta}{\hat{r}}&= {\cal E}\,\hat{T}^{\,3/2}.\label{e39}
\end{align}
Note that ${\cal E}(t)$ can be adjusted by changing the time derivative of the  current flowing in the central solenoid.  Furthermore, tokamaks
invariably  possess a feedback system that automatically alters the instantaneous value of ${\cal E}(t)$ in such a manner as to produce a prescribed plasma
current waveform, $I_p(t)$. 

\subsection{Rescaling}
Suppose that the instantaneous minor radius of the plasma is $\delta(\hat{t})\,a$. In other words, suppose that the limiter
is situated at $\hat{r}= \delta$. Let $x=\hat{r}/\delta$, and let
\begin{equation}
\hat{T}(\hat{r},\hat{t}) = \frac{\lambda^2}{(\delta\,{\cal E})^4}\,\bar{T}(x).
\end{equation}
Equation~(\ref{e38}) transforms to give
\begin{equation}\label{e41}
\frac{d^2\bar{T}}{dx^2} + \frac{1}{x}\,\frac{d\bar{T}}{dx} + \frac{d\ln\hat{\chi}}{dx}\,\frac{d\bar{T}}{dx} + \frac{\lambda\,\bar{T}^{\,3/2}}{\hat{\chi}}=0.
\end{equation}
Appropriate boundary conditions are 
\begin{align}
\bar{T}(0) &=1,\\[0.5ex]
\frac{d\bar{T}(0)}{dx}& = 0,\\[0.5ex]
\bar{T}(1) &= 0.\label{bc}
\end{align}
The parameter $\lambda$ must be adjusted until Eq.~(\ref{bc}) is satisfied. Clearly, the determination of the normalized electron temperature profile, $\bar{T}(x)$, 
involves the solution of a nonlinear eigenvalue equation, (\ref{e41}), where $\lambda$ plays the role of the eigenvalue. Of course, the electron temperature is not zero at the
plasma boundary, but is, instead, determined by sheath physics at the plasma/limiter interface. However, we expect the edge temperature to
be much lower than that in the plasma interior, which justifies the approximation $\bar{T}(1)=0$. 

Let 
\begin{equation}
\hat{B_\theta}(\hat{r},\hat{t}) = \frac{\lambda^3}{(\delta\,{\cal E})^5}\,\bar{B}_\theta(x).
\end{equation}
Equation~(\ref{e39}) yields 
\begin{equation}
\frac{d\bar{B}_\theta}{dx} + \frac{\bar{B}_\theta}{x} = \bar{T}^{\,3/2},
\end{equation}
which must be solved subject to the boundary condition
\begin{equation}
\frac{d\bar{B}(0)}{dx}=0.
\end{equation}

Let
\begin{equation}
q(\hat{r},\hat{t}) =\frac{q_a\,\delta\,(\delta\,{\cal E})^5}{\lambda^3}\,\bar{q}(x).
\end{equation}
Equation~(\ref{e28}) gives
\begin{equation}
\bar{q}(x) = \frac{x}{\bar{B}_\theta(x)}.
\end{equation}
Note that
\begin{equation}
\bar{q}(0)= 2.
\end{equation}
According to Eq.~(\ref{e29}), the net normalized toroidal current flowing in the plasma is
\begin{equation}
\hat{I}_p(\hat{t}) \equiv \hat{I}_p(\delta,\hat{t})= \frac{\lambda^3\,\delta}{(\delta\,{\cal E})^5}\,\bar{B}_\theta(1).
\end{equation}
Equation~(\ref{e36}) implies that the normalized ohmic heating power is
\begin{equation}
\hat{P}(\hat{t})= \frac{\lambda^2}{(\delta\,{\cal E})^4}\,\zeta,
\end{equation}
where
\begin{equation}
\zeta = - \left(\hat{\chi}\,x\,\frac{d\bar{T}}{dx}\right)_{x=1}.
\end{equation}
According to Eq.~(\ref{e30}), the total normalized poloidal magnetic flux contained between the magnetic axis and the plasma
boundary is
\begin{equation}
\hat{\psi}_{\rm tot}(\hat{t}) = \frac{\lambda^{\,3}\,\delta}{(\delta\,{\cal E})^5}\,\int_0^1\bar{B}_\theta\,dx.
\end{equation}

Finally, the criterion (\ref{e35}) becomes 
\begin{equation}\label{e55}
{\cal E}> \delta\left|\frac{d}{d\hat{t}}\!\left[\frac{\lambda^3}{(\delta\,{\cal E})^5}\right]\right|\int_0^1\bar{B}(x)\,dx.
\end{equation}
Note that we have imposed the criterion at $r=\delta\,a$, rather than $r=a$, because it only needs to be satisfied
within the plasma. 

\section{Current Ramp Scenarios}
\subsection{Current Ramp-Up}
\subsubsection{Current Waveform}
Suppose that the current ramp-up lasts from $t=0$ to $t=t_{\rm ramp}$. Suppose that, during the ramp-up, the current flowing in the central solenoid is adjusted such that
the normalized plasma current increases {\em linearly}\/ in time:
\begin{equation}\label{e56}
\hat{I}_p(\hat{t}) = \frac{\hat{t}}{\hat{t}_{\rm ramp}},
\end{equation}
where $\hat{t}_{\rm ramp} =t_{\rm ramp}/\tau_R$. For the sake of simplicity, we shall assume that the plasma elongation, $\kappa$,  is constant in time. 

\subsubsection{Initial Stage}\label{init}
Suppose that, during the initial stage of the ramp-up, which lasts from $t=0$ to $t=t_{\rm pre}$, where $t_{\rm pre}\ll t_{\rm ramp}$, 
the safety-factor at the plasma boundary is allowed to decrease from an initial very large value to its final value, $q_a$, in such a manner that
\begin{equation}\label{e57}
q(x,\hat{t}) = q_a\left(\frac{\hat{t}_{\rm pre}}{\hat{t}}\right)\frac{\bar{q}(x)}{\bar{q}(1)},
\end{equation}
where $\hat{t}_{\rm pre}= t_{\rm pre}/\tau_R$. Here, $q_a$ is chosen such that $q(0)=1$ at the end of the initial stage. We make this choice because the
sawtooth oscillation effectively prevents the central safety-factor from falling significantly below unity.\cite{book} It follows that
\begin{equation}
q_a = \frac{1}{2\,\bar{B}_\theta(1)}.
\end{equation}
Equations~(\ref{e56}) and (\ref{e57}) allow us to determine all other quantities in the initial
stage of the ramp-up. We obtain
\begin{equation}
\delta = \left(\frac{\hat{t}_{\rm pre}}{\hat{t}_{\rm ramp}}\right)^{1/2},
\end{equation}
which implies that the plasma minor radius takes a relatively small constant value during the first stage of the ramp-up. 
Furthermore,
\begin{align}
\hat{T}(x,\hat{t}) &= T_{\rm ramp}\left(\frac{\hat{t}_{\rm pre}}{\hat{t}_{\rm ramp}}\right)^{2/5}\left(\frac{\hat{t}}{\hat{t}_{\rm pre}}\right)^{4/5}\bar{T}(x),\\[0.5ex]
\hat{B}_\theta(x,\hat{t})&= \left(\frac{\hat{t}_{\rm pre}}{\hat{t}_{\rm ramp}}\right)^{1/2}\left(\frac{\hat{t}}{\hat{t}_{\rm pre}}\right)\frac{\bar{B}_\theta(x)}{\bar{B}_\theta(1)},\\[0.5ex]
{\cal E}(\hat{t}) &= {\cal E}_{\rm ramp}\left(\frac{\hat{t}_{\rm pre}}{\hat{t}_{\rm ramp}}\right)^{-3/5}\left(\frac{\hat{t}}{\hat{t}_{\rm pre}}\right)^{-1/5},\\[0.5ex]
\hat{P}(\hat{t}) &= P_{\rm ramp}\left(\frac{\hat{t}_{\rm pre}}{\hat{t}_{\rm ramp}}\right)^{2/5}\left(\frac{\hat{t}}{\hat{t}_{\rm pre}}\right)^{4/5},\\[0.5ex]
\hat{\psi}_{\rm tot}(\hat{t})&= \left(\frac{\hat{t}}{\hat{t}_{\rm ramp}}\right)\int_0^1\frac{\bar{B}_{\theta}(x)}{\bar{B}_\theta(1)}\,dx,
\end{align}
where
\begin{align}
T_{\rm ramp}&= \lambda^{-2/5}\,[\bar{B}_\theta(1)]^{-4/5},\\[0.5ex]
{\cal E}_{\rm ramp} &= \lambda^{3/5}\,[\bar{B}_\theta(1)]^{1/5},\\[0.5ex]
P_{\rm ramp} &= \zeta\,T_{\rm ramp}.
\end{align}
Finally, the criterion (\ref{e55}) yields
\begin{equation}\label{e55d}
\hat{t}_{\rm ramp} > \left(\frac{\hat{t}_{\rm pre}}{\hat{t}_{\rm ramp}}\right)^{3/5}\left(\frac{\hat{t}}{\hat{t}_{\rm pre}}\right)^{1/5}{\cal E}_{\rm ramp}^{-1}\int_0^1\frac{\bar{B}_{\theta}(x)}{\bar{B}_\theta(1)}\,dx,
\end{equation}
which does indeed constitute a limitation on the maximum current ramp rate.

\subsubsection{Main Stage}\label{main}
In the main stage of the current ramp-up, the plasma current still increases linearly in time according to Eq.~(\ref{e56}), but the edge safety-factor value
is held constant in such a manner that 
\begin{equation}\label{e57a}
q(x,\hat{t}) = q_a\,\frac{\bar{q}(x)}{\bar{q}(1)}.
\end{equation}
As before, Eqs.~(\ref{e56}) and (\ref{e57a}) allow us to determine all other quantities in the main 
stage of the ramp-up. We find that 
\begin{align}
\delta &= \left(\frac{\hat{t}}{\hat{t}_{\rm ramp}}\right)^{1/2},\label{e70}\\[0.5ex]
\hat{T}(x,\hat{t}) &= T_{\rm ramp}\left(\frac{\hat{t}}{\hat{t}_{\rm ramp}}\right)^{2/5}\bar{T}(x),\\[0.5ex]
\hat{B}_\theta(x,\hat{t})&= \left(\frac{\hat{t}}{\hat{t}_{\rm ramp}}\right)^{1/2}\frac{\bar{B}_\theta(x)}{\bar{B}_\theta(1)},\\[0.5ex]
{\cal E}(\hat{t}) &= {\cal E}_{\rm ramp}\left(\frac{\hat{t}}{\hat{t}_{\rm ramp}}\right)^{-3/5},\\[0.5ex]
\hat{P}(\hat{t}) &= P_{\rm ramp}\left(\frac{\hat{t}}{\hat{t}_{\rm ramp}}\right)^{2/5},\\[0.5ex]
\hat{\psi}_{\rm tot}(\hat{t})&= \left(\frac{\hat{t}}{\hat{t}_{\rm ramp}}\right)\int_0^1\frac{\bar{B}_{\theta}(x)}{\bar{B}_\theta(1)}\,dx.
\end{align}
It is clear from Eq.~(\ref{e70}) that, in order to keep the edge safety-factor constant, while the plasma current increases in time, it
is necessary for the minor radius of the plasma to also increase in time. 

The criterion (\ref{e55}) yields
\begin{equation}
\hat{t}_{\rm ramp} > \left(\frac{\hat{t}}{\hat{t}_{\rm ramp}}\right)^{3/5}\frac{1}{2\,{\cal E}_{\rm ramp}}\int_0^1\frac{\bar{B}_{\theta}(x)}{\bar{B}_\theta(1)}\,dx,
\end{equation}
Now, if this criterion is satisfied at the end of the current ramp then it is satisfied at all times in the main current ramp, and the criterion (\ref{e55d})
is also satisfied at all times in the initial stage of the current ramp. 
We conclude that the criterion (\ref{e55}) is satisfied at all times in the current ramp provided 
\begin{equation}
t_{\rm ramp} > \tau_{\rm min},
\end{equation}
where \begin{equation}\label{taum}
\tau_{\rm min}= \frac{\tau_R}{2\,{\cal E}_{\rm ramp}}\int_0^1\frac{\bar{B}_{\theta}(x)}{\bar{B}_\theta(1)}\,dx,
\end{equation}
Clearly, $\tau_{\rm min}$ represents the minimum possible completely safe current ramp-up timescale. 

\subsection{Current Ramp-Down}
Suppose that the current ramp-down lasts from $t=0$ to $t=t_{\rm ramp}$. Suppose that, during the ramp-down, 
the normalized plasma current decreases linearly in time:
\begin{equation}
\hat{I}_p(\hat{t}) = \frac{\hat{t}_{\rm ramp}-\hat{t}}{\hat{t}_{\rm ramp}}.
\end{equation}
In the main stage of the ramp-down, the edge safety-factor is held constant, so that Eq.~(\ref{e57a}) is valid.
In the final stage of the ramp-down, which lasts from $t=t_{\rm ramp}-t_{\rm pre}$ to $t=t_{\rm ramp}$,
the edge safety-factor increases in such a manner that 
\begin{equation}
q(x,\hat{t}) = q_a\left(\frac{\hat{t}_{\rm pre}}{\hat{t}_{\rm ramp}-\hat{t}}\right)\frac{\bar{q}(x)}{\bar{q}(1)}.
\end{equation}
The time variation of other quantities during the main stage of ramp-down can be obtained by making the transformation
$\hat{t}\rightarrow \hat{t}_{\rm ramp} - \hat{t}$ in the relevant formulae  given in Sect.~\ref{main}. Likewise,
the time variation of other quantities during the final stage of the ramp down are obtained be making the
same transformation in the relevant formulae given in Sect.~\ref{init}. Clearly, the current ramp-down is the mirror image in time of the current ramp-up.
In particular, the minimum completely safe current
ramp-down timescale, $\tau_{\rm min}$, is specified in Eq.~(\ref{taum}). 

\section{Results}
\subsection{Plasma Equilibrium}
Let our normalized perpendicular energy diffusivity profile take the form
\begin{equation}
\hat{\chi}(x) = f(\alpha)\,(1+x^2)^\alpha,
\end{equation}
where
\begin{equation}
f(\alpha) = \frac{1+\alpha}{2^{\,1+\alpha}-1}.
\end{equation}
Note that $\int_0^1\,\hat{\chi}(x)\,dx/\int_0^1x\,dx=1$, which ensures that $\chi_0$ is the volume averaged perpendicular diffusivity, in accordance with our earlier assumption. 

\begin{figure}
\centerline{\includegraphics[width=0.85\textwidth]{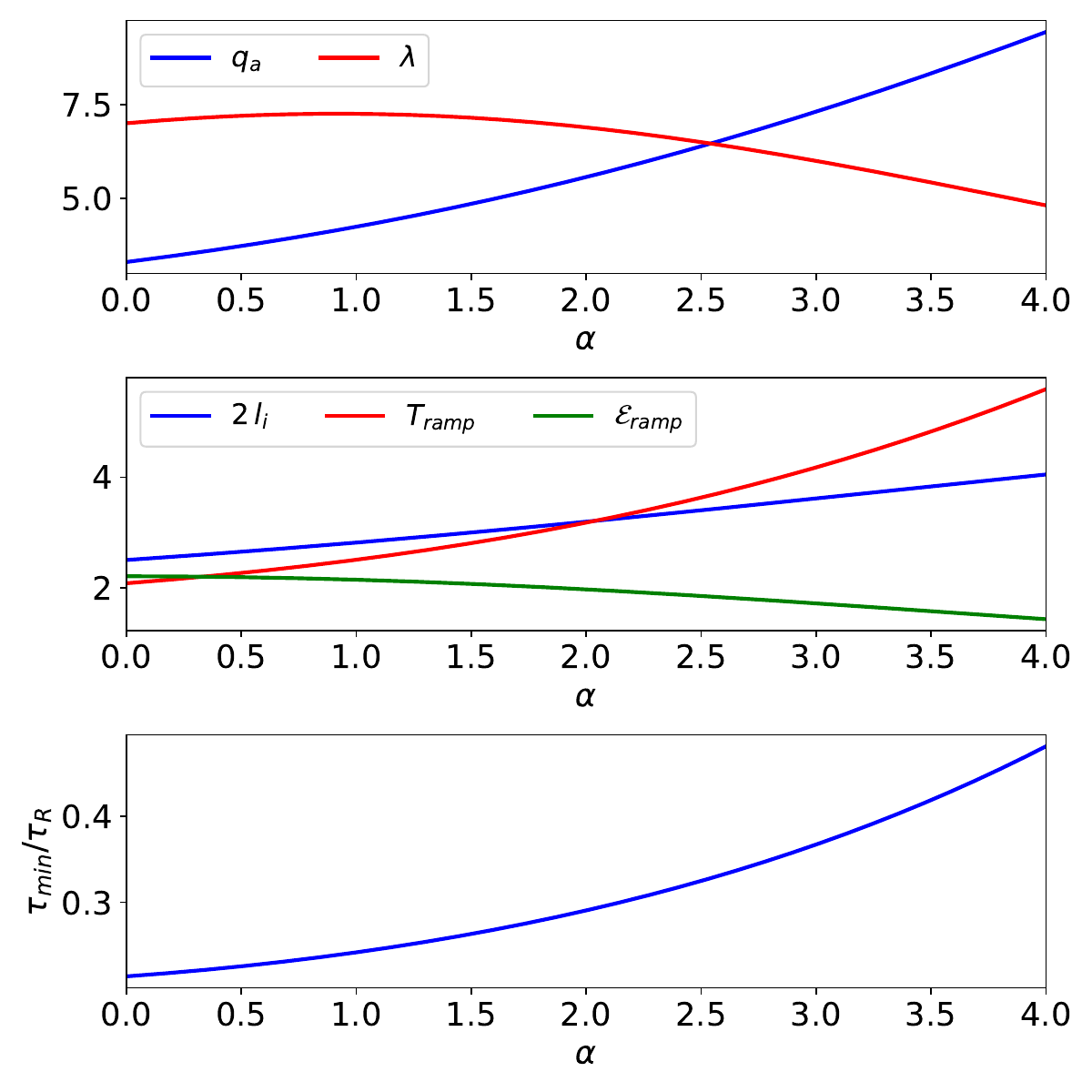}}
\caption{Various equilibrium quantities calculated as functions of the diffusivity profile parameter, $\alpha$.
The top panel shows the eigenvalue, $\lambda$, and the edge safety-factor value, $q_a$. 
The middle panel shows the normalized plasma self-inductance, $l_i$, the normalized central electron temperature, $T_{\rm ramp}$,
and the normalized inductive electric field, ${\cal E}_{\rm ramp}$. The lower panel shows the ratio of the minimum safe current ramp
timescale, $\tau_{\rm min}$, to the resistive diffusion timescale, $\tau_R$. }\label{fig2}
\end{figure}

Figure~\ref{fig2} shows various equilibrium quantities plotted as a function of $\alpha$. As $\alpha$ increases, which implies that the diffusivity at the plasma
boundary becomes larger than that in the core, the equilibrium current profile becomes more peaked. This is clear from the figure because both the edge safety-factor
value, $q_a$, and the normalized internal plasma inductance,\cite{cheng}
\begin{equation}
l_i= \frac{2}{\bar{B}_\theta^{\,2}(1)}\int_0^1\bar{B}_\theta^{\,2}(x)\,x\,dx,
\end{equation}
increase with increasing $\alpha$. (A peaked current profile is associated with  large $q_a/q_0$ and $l_i$ values, and vice versa.\cite{wesson,book}) The peaking of the current profile also causes a decrease in the normalized central electric field, ${\cal E}_{\rm ramp}$,  with increasing $\alpha$. 
Note that $P_{\rm ramp}= {\cal E}_{\rm ramp}$ for the set of equilibria in question. Finally, the minimum safe current ramp timescale, $\tau_{\rm min}$,  is an increasing function of $\alpha$. 

\begin{figure}
\centerline{\includegraphics[width=0.7\textwidth]{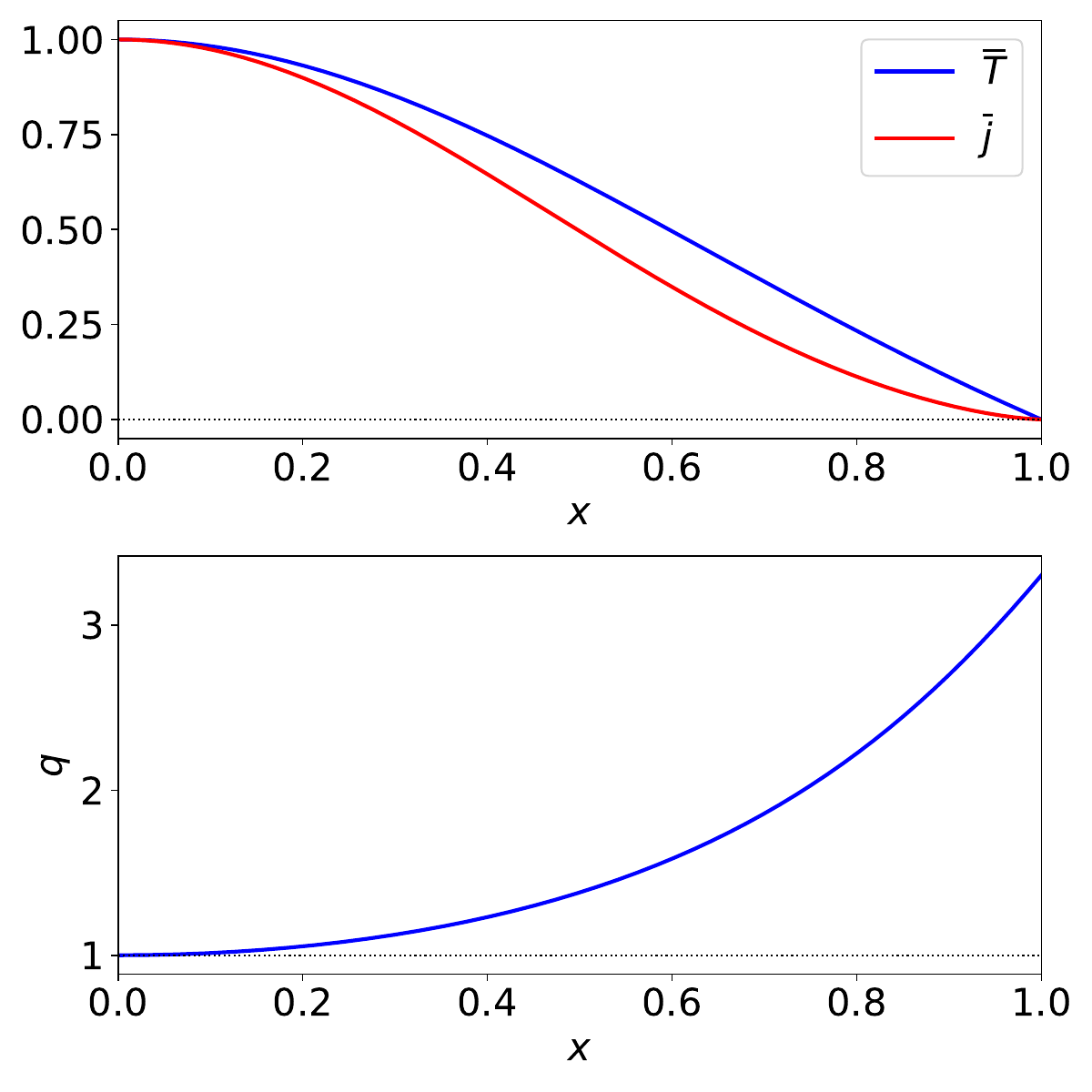}}
\caption{Equilibrium profiles calculated as functions of the normalized minor radius, $x$, for $\alpha=0$.  The top panel shows the normalized electron temperature profile, $\bar{T}(x)$, and the
normalized current profile, $\bar{j}(x)= \bar{T}^{\,3/2}$.  The bottom panel shows the safety-factor profile, $q(x)= q_a\,\bar{q}(x)$. }\label{fig3}
\end{figure}

For the sake of simplicity, let us consider the case $\alpha=0$, in which the perpendicular diffusivity is uniform within the plasma. In this situation,
we find that 
$\lambda =  7.01$,  $q_a    =  3.30$,   $l_i   =  1.25$, 
$T_{\rm ramp} =  2.08$,  ${\cal E}_{\rm ramp}=P_{\rm ramp}  =  2.21$, and $\tau_{\rm min}/\tau_R =  0.214$. 
Figure~\ref{fig3} shows the corresponding normalized electron temperature and current profiles, as well as the safety-factor profile. 
It can be seen that the current profile is fairly broad. Nevertheless, the equilibrium lies in the stable region of the
conventional $q_a$--$l_i$ diagram. (See Fig.~4 of Ref.~\onlinecite{cheng}.) Note that, due to profile effects,  the minimum ramp timescale,
$\tau_{\rm min}$,  is about one fifth of the conventional
resistive diffusion timescale, $\tau_R$. 

\subsection{Simulated Current Ramp-Up}
In this section, we shall describe simulated current ramp-ups made using our model for the JET, SPARC, and ITER tokamaks. The
machine parameters for these tokamaks are given in Table~\ref{t1}. The simulations are made assuming that $\chi_0=1\,{\rm m\,s}^{-1}$,
$\ln{\mit\Lambda} = 15$, $Z=2$, $t_{\rm pre}= 0.05\,t_{\rm ramp}$, and $t_{\rm ramp} =\tau_{\rm min}$. Note that the current ramp time in these simulations takes its
minimum completely safe value, according to our model. 

\begin{figure}
\centerline{\includegraphics[width=0.9\textwidth]{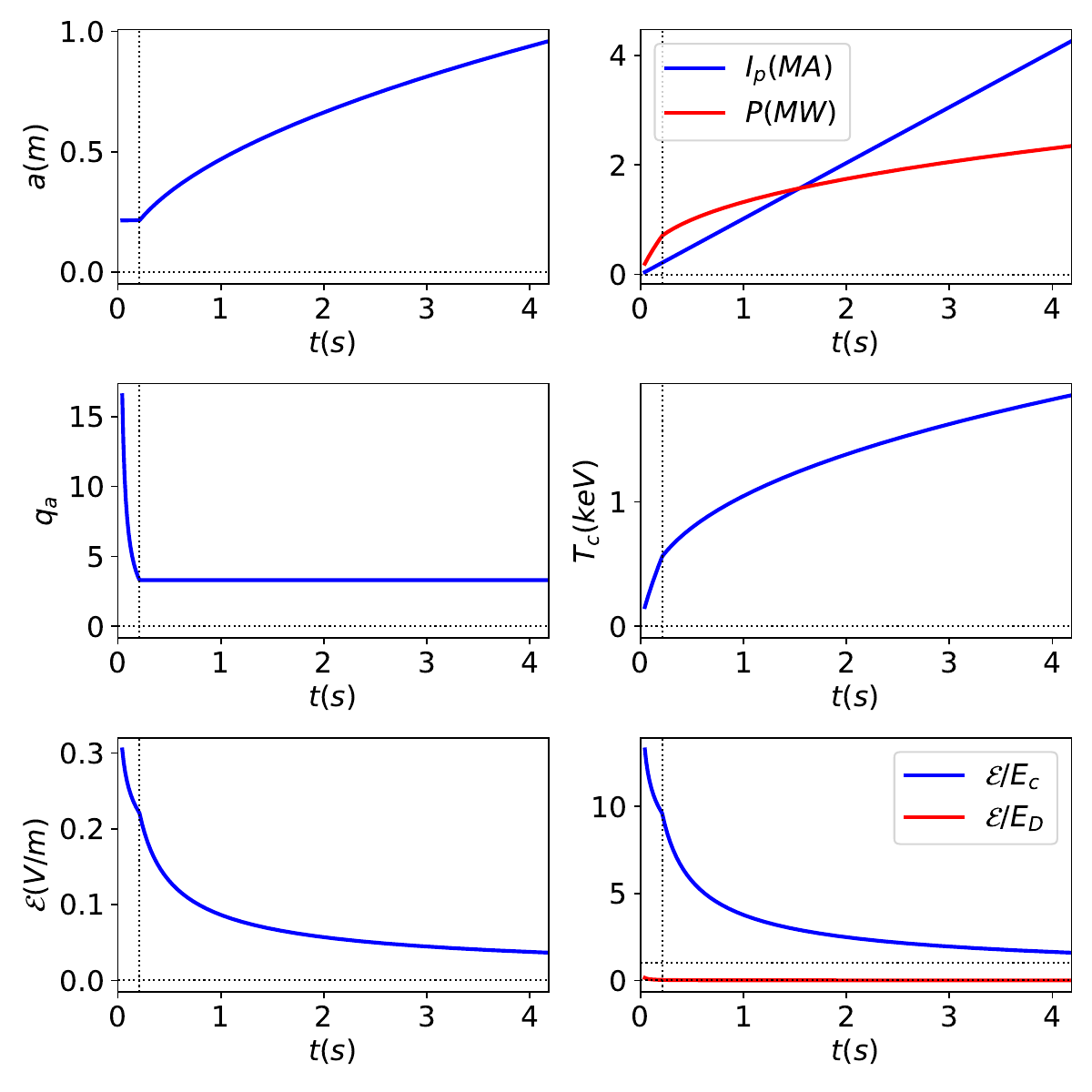}}
\caption{Simulated current ramp-up for the JET tokamak. Here, $a$ is the minor radius of the plasma, $I_p$ the toroidal plasma
current, $P$ the ohmic power, $q_a$ the edge safety-factor, $T_c$ the central electron temperature, ${\cal E}$ the
inductive electric field, $E_c$ the Connor-Hastie field, and $E_D$ the central Dreicer field. }\label{fig4}
\end{figure}

Figure~\ref{fig4} shows a simulated current ramp-up for JET. Some of the characteristic  parameters for this ramp-up are listed in Table~\ref{t2}. 
It can be seen that the toroidal plasma current in JET can be safely ramped up to about 4.3 MA in about 4.2 seconds. This prediction accords well
with operational experience in JET. (See Fig.~1 of Ref.~\onlinecite{jet}.)
Note that, as reported by de Vries et alia (2020),\cite{run} the inductive electric field exceeds the Connor-Hastie field during the current ramp-up. However,
the electric field only exceeds de Vries et alia's (2020) empirical runaway electron generation threshold, which is ten times the Connor-Hastie field,\cite{run} in
the very early stages of the current ramp-up. Note that the threshold is exceeded, not because the current ramp  rate is too large, but rather because the
plasma is very cold and resistive in the early stages of the ramp-up. One obvious solution to runaway electron problems in the early
stages of a current ramp-up is to apply small amounts of electron cyclotron wave heating in order to slightly raise the electron temperature. 
The inductive electric field is slightly above the Connor-Hastie value at the end of the current ramp-up, but we would expect the field to
fall as soon as the main external heating is applied. Hence, the inductive electric field should be well below the Connor-Hastie
value throughout most of the current flat-top. Note that the inductive electric field is very much less than the Dreicer field throughout the current ramp-up, as is also
the case for our SPARC and ITER simulations. 

\begin{figure}
\centerline{\includegraphics[width=0.9\textwidth]{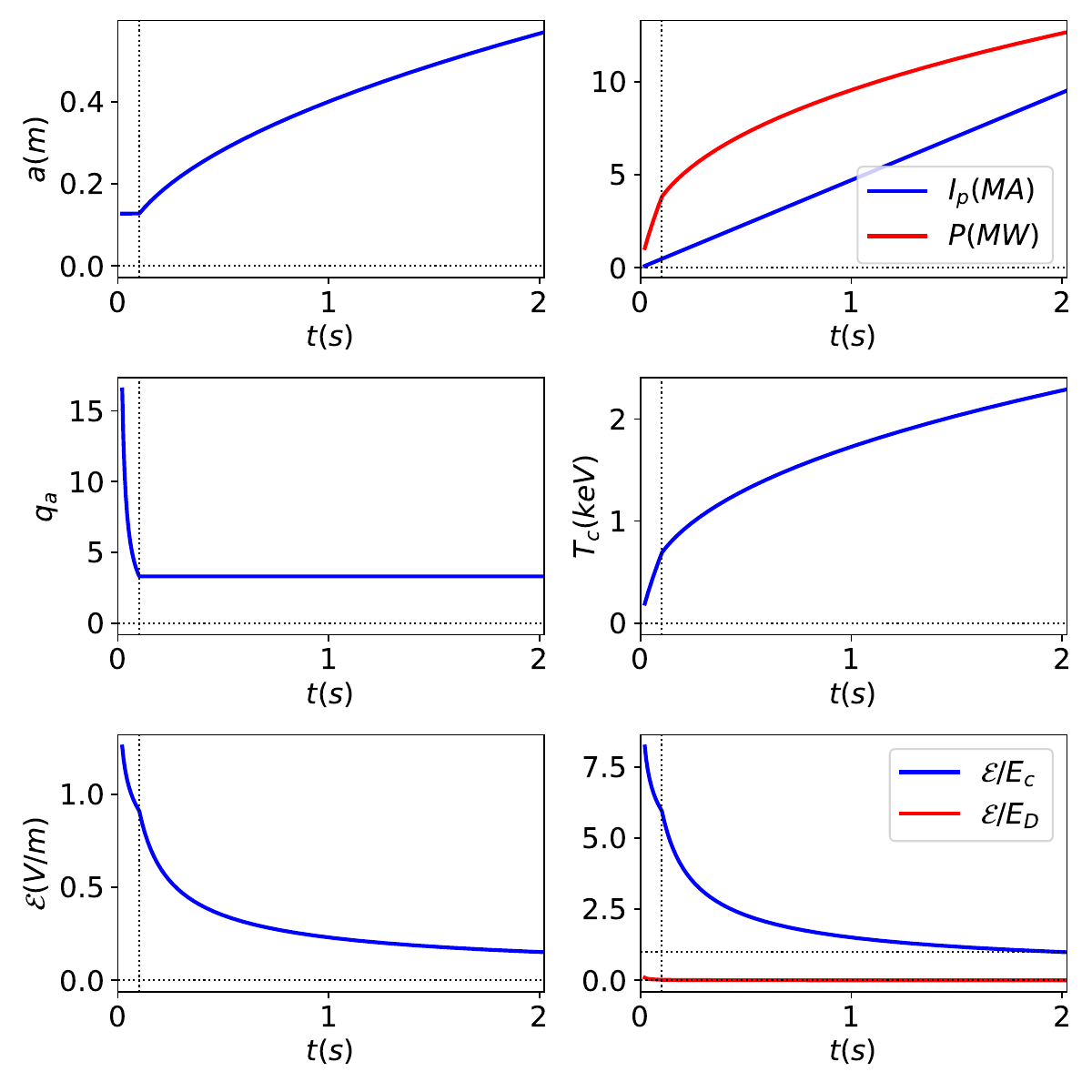}}
\caption{Simulated current ramp-up for the SPARC tokamak. See caption to Fig.~\ref{fig4}.}\label{fig5}
\end{figure}

Figure~\ref{fig5} shows a simulated current ramp-up for SPARC. Some of the characteristic  parameters for this ramp-up are listed in Table~\ref{t2}. 
It can be seen that the toroidal plasma current in SPARC can be safely ramped up to about 9.5 MA in about 2.0 seconds. This maximum safe ramp-rate is 
actually greater than the ramp-rate in the SPARC design.\cite{creely} 
Note that the inductive electric field, ${\cal E}$,  exceeds the Connor-Hastie field, $E_c$,  during the current ramp-up. However, the ratio ${\cal E}/E_c$ for SPARC,
shown  in Fig.~\ref{fig5}, is less than that the corresponding ratio for JET,  shown in Fig.~\ref{fig4}. In other words, if JET was capable of
operating without dangerous levels of runaway electron generation during the current ramp-up, as reported by  Vries et alia (2020),\cite{run} then SPARC
should certainly also be capable of operating without runaway electron problems. 

\begin{figure}
\centerline{\includegraphics[width=0.9\textwidth]{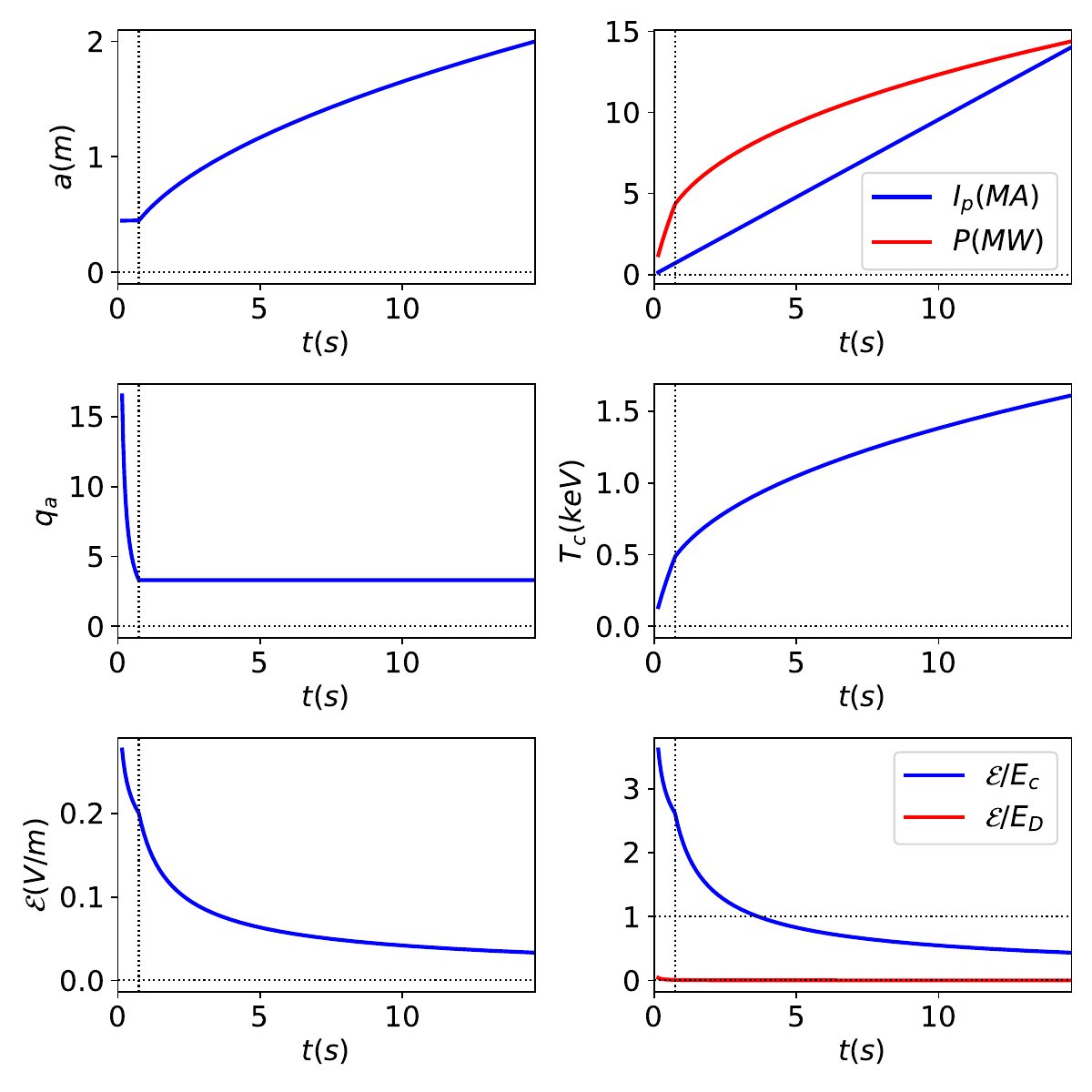}}
\caption{Simulated current ramp-up for the ITER tokamak. See caption to Fig.~\ref{fig4}.}\label{fig6}
\end{figure}

Finally, Fig.~\ref{fig6} shows a simulated current ramp-up for ITER. Some of the characteristic  parameters for this ramp-up are listed in Table~\ref{t2}. 
It can be seen that the toroidal plasma current in ITER can be safely ramped up to about 14.0 MA in about 14.7 seconds. This maximum safe ramp-rate is 
consistent with the series of equilibria shown in Fig.~1 of Ref.~\onlinecite{federici}, and is much greater than the 15 MA in 60 s ramp-down rate envisioned by de Vries et alia (2018).\cite{deVries}
Note that the inductive electric field, ${\cal E}$,  exceeds the Connor-Hastie field, $E_c$,  during the first third of the current ramp-up. However,  the ratio ${\cal E}/E_c$ for ITER, 
shown  in Fig.~\ref{fig6}, is very much less than that the corresponding ratio for JET,  shown in Fig.~\ref{fig4}. As before, we conclude that if JET was capable of
operating without dangerous levels of runaway electron generation during the current ramp-up, then ITER
should certainly also be capable operating without runaway electron problems.  

\section{Summary}
A simple model of the ramp-up and ramp-down of the toroidal current in a tokamak plasma is developed. Faraday's law of electric induction is found to 
limit how rapidly the current can be safety ramped up or down. We estimate that the minimum completely safe ramp-up/down times for the JET, SPARC, and
ITER tokamaks are 4.2, 2.0, and 14.7 seconds, respectively. The JET ramp time is in accordance with operational experience. The SPARC
and ITER minimum safe ramp times are less than the ramp times in the respective designs. Hence, there is no indication that the design ramp times are infeasible, as was recently 
suggested by Boozer.\cite{boozer}
We also find that the typical ratios of the inductive electric field
to the Connor-Hastie field in SPARC and ITER are less than those in JET. Thus, the fact that the JET tokamak was able to operate successfully
without encountering runaway electron problems during  current ramps suggests that the future SPARC and ITER tokamaks should certainly be able to
avoid such problems. 

\section*{Acknowledgements}
This research was funded by the  U.S.\ Department of Energy, Office of Science, Office of Fusion Energy Sciences.

\section*{Data Availability Statement}
The digital data used in the figures in this paper can be obtained from the author upon reasonable request.

\newpage 
\begin{table}
\begin{tabular}{cccccc}\hline
Machine & ~$R_0({\rm m})$~ & ~$a({\rm m})$~ & ~$B_0({\rm T})$~ & ~$n_e(10^{20}\,{\rm m}^{-3})$~ & ~$\kappa$\\ \hline
JET & 2.96 & 0.96 & 3.45 & 0.3& 1.7\\[0.5ex]
SPARC & 1.85 & 0.57 & 12.2 & 2.0 & 1.97\\[0.5ex]
ITER & 6.2 & 2.0 & 5.3 & 1.0 & 1.85
\end{tabular}
\caption{Machine parameters for the JET, SPARC, and ITER tokamaks. 
The JET parameters come from Ref.~\onlinecite{jetx}. 
The SPARC and ITER parameters are taken from Tables 1 and 2 of Ref.~\onlinecite{creely}, as well as Table~1 of Ref.~\onlinecite{iter}. 
Here, $R_0$ is the major radius, $a$ the minor radius, $B_0$ the toroidal magnetic field-strength, $n_e$ the electron number
density, and $\kappa$ the vertical elongation.}\label{t1}
\end{table}

\begin{table}
\begin{tabular}{ccccccccc}\hline
Machine & ~$I_p({\rm MA})$~ & ~$P({\rm MW})$~ & ~$T_c({\rm keV})$~ &~$\beta_p$~& ~${\cal E}({\rm V\,m}^{-1})$~ & ~$E_c({\rm V\,m}^{-1})$~ & ~$E_D({\rm V\,m}^{-1})$~& ~$\tau_{\rm min}({\rm s)}$\\ \hline
JET & 4.26 & 2.34& 1.86  &0.0470& 0.0336&0.0229&13.1 & 4.19\\[0.5ex]
SPARC & 9.53& 12.7 & 2.29  &0.0343& 0.151&0.153&70.9& 2.02\\[0.5ex]
ITER & 14.0 & 14.4 & 1.61  &0.0583& 0.0332&0.0765& 50.4&14.7
\end{tabular}
\caption{Plasma parameters at end of current ramp for the JET, SPARC, and ITER tokamaks. Here, $I_p$ is the toroidal plasma current, $P$ the
ohmic heating power, $T_c$ the central electron temperature, $\beta_p$ the central poloidal beta, ${\cal E}$ the inductive
electric field, $E_c$ the Connor-Hastie field, and $E_D$ the central Dreicer field. Finally, $\tau_{\rm min}$ is the minimum safe current ramp timescale.  }\label{t2}
\end{table}

\begin{thebibliography}{99}\baselineskip 5ex

\bibitem{boozer}   A.H.~Boozer, arXiv preprint arXiv:2507.05456 (2025). 

\bibitem{deVries} P.C.~de Vries, T.C.~Luce, Y.S.~Bae, S.~Gerhardt, X.~Gong, Y.~Gribov, 
D.~Humphreys, A.~Kavin, R.R.~Khayrutdinov, C.~Kessel, et al., Nucl.\ Fusion {\bf 58} 026019 (2018).

\bibitem{boozer1} A.H.~Boozer, Nucl.\ Fusion {\bf 61}, 054004 (2021). 

\bibitem{creely} A.J.~Creely, M.J.~Greenwald, S.B.~Ballinger, D.~Brunner, J.~Canik, J.~Dooly, 
T.~F\"{u}lop, D.T.~Garnier, R.~Granetz, et al.,  J.\ Plasma Phys.\ {\bf 86}, 865860502 (2020).

\bibitem{corsica} T.~Casper, Y.~Gribov, A.~Kavin, V.~Lukash, R.R.~Khayrutdinov, H.~Fujieda and C.~Kessel, Nucl.\ Fusion {\bf 54}, 013005 (2014).

\bibitem{corsica1} S.H.~Kim, R.H.~Bulmer, D.J.~Campbell, T.A.~Casper, L.L.~LoDestro, W.H.~Meyer, L.D.~Pearlstein and J.A.~Snipes,
Nucl.\ Fusion {\bf 56}, 126002 (2016).

\bibitem{tsc} S.C.~Jardin, N.~Pomphrey and J.~Delucia, J.\ Comput.\ Phys.\ {\bf 66}, 481 (1986).

\bibitem{lister} J.B.~Lister, A.~Portone and Y.~Gribov, IEEE Control Systems Magazine {\bf 26}, no. 2, 79 (2006).

\bibitem{federici} G.~Federici, O.~Zolotukhin, M.~Kobayashi, A.~Loarte, G.~Strohmayer, A.~Tanga, A.~Portone, L.~Horton,
Y.~Feng, F.~Sardei, et al., J.\ Nucl.\ Materials {\bf 363}, 346 (2007). 

\bibitem{jackson} G.L.~Jackson, P.A.~Politzer, D.A.~Humphreys, T.A.~Casper, A.W.~Hyatt, J.A.~Leuer, J.~Lohr,
T.C.~Luce, M.A.~Van Zeeland and J.H.~Yu, Phys.\ Plasmas {\bf 17}, 056116 (2010).

\bibitem{politzer} P.A.~Politzer, G.L.~Jackson, D.A.~Humphreys, T.C.~Luce, A.W.~Hyatt and J.A.~Leuer, Nucl.\ Fusion {\bf 50}, 035011 (2010).

\bibitem{wesson} J.A.~Wesson, Nucl.\ Fusion {\bf 18}, 87 (1987).

\bibitem{granetz1} R.S.~Granetz, I.J.~Hutchinson and D.O.~Overskei, Nucl.\ Fusion {\bf 19}, 1587 (1979).

\bibitem{cheng} C.Z.~Cheng, H.P.~Furth and A.H.~Boozer, Plasma Phys.\ Control.\ Fusion {\bf 29}, 351 (1987).

\bibitem{uckam} N.A.~Uckan and the ITER Physics Group, {\em ITER Physics Design Guidelines: 1989},  (IAEA, Vienna, 1990).

\bibitem{spitzer} L.~Spitzer, Jr., {\em Physics of Fully Ionized Gases}\/ (Interscience, New York NY, 1956).

\bibitem{fitz} R.~Fitzpatrick, {\em Plasma Physics: An Introduction}, 2nd ed., (CRC, Boca Raton FL, 2023).

\bibitem{fitz1} R.~Fitzpatrick, {\em Tearing Mode Dynamics in Tokamak Plasmas}, (IOP, Bristol UK, 2023).

\bibitem{book} J.A.~Wesson, {\em Tokamaks}, 4th ed., (Oxford, Oxford UK, 2011).

\bibitem{connor} J.W.~Connor and R.J.~Hastie, Nucl.\ Fusion {\bf 15}, 415 (1975).

\bibitem{dreicer} H.~Dreicer, Phys.\ Rev.\ {\bf 115}, 238 (1959).

\bibitem{run} P.C.~de\,Vries, Y.~Gribov, J.R.~Martin-Solis, A.B.~Mineev, J.~Sinha, A.C.C.~Sips, V.~Kiptily, A.~Loarte and JET Contributors, Plasma Phys.\ Control.\ Fusion
{\bf 62}, 125014 (2020). 

\bibitem{granetz} R.~Granetz, B.~Eposito, J.H.~Kim, R.~Koslowski, M.~Lehnen, J.R.~Martin-Solis, C.~Paz-Solden, T.~Rhee, J.C.~Wesley
and L.~Zheng, Phys.\ Plasmas {\bf 21}, 072506 (2014).

\bibitem{paz} C.~Paz-Solden,  N.W.~Eidietis,  R.~Granetz, E.M.~Hollmann, R.A.~Moyer, J.C.~Wesley, J.~Zhang, M.E.~Austin, N.A.~Crocker, A.~Wingen and Y.~Zhu, 
 Phys.\ Plasmas {\bf 21}, 022514 (2014).

\bibitem{pop} Z.~Popovic, B.~Esposito, J.R.~Martin-Solis, W.~Bin,  P.~Buratti, D.~Carnevale, F.~Causa, M.~Gospodarczyk,
D.~Marocco, G.~Ramogida and M.~Riva, Phys.\ Plasmas {\bf 23}, 122501 (2016). 

\bibitem{jet} I.~Voitsekhovitch, A.C.C.~Sips, B.~Alper, M.~Beurskens, I.~Coffey,
J.~Conboy, T.~Gerbaud, C.~Giroud1, T.~Johnson, F.~K\"{o}ch, et al, Plasma Phys.\ Control. Fusion {\bf 52}  105011 (2010). 

\bibitem{jetx} Z.S.~Hartwig and Y.A.~Podpaly, {\em Magnetic Fusion Energy Formulary}, (MIT, Cambridge MA, 2011).
 
\bibitem{iter} ITER Physics Basis Editors, et al., Nucl.\ Fusion {\bf 39} 2137 (1999). 
 
\end{thebibliography}
\end{document}